\documentclass[12pt,english]{iopart}
\usepackage{graphicx}
\usepackage{amssymb}

\makeatletter
\usepackage{iopams}
\usepackage{setstack}

\usepackage{iopams}  
\makeatother

\usepackage{babel}

\begin{document}

\title{A structural comparison of models of colloid-polymer mixtures}

\author{Jade Taffs$^{1}$, Alex Malins$^{2}$, Stephen R. Williams$^{3}$
 and C. Patrick Royall$^{1}$ }

\address{$^{1}$ School of Chemistry, University of Bristol, Bristol, BS8
1TS, UK}

\address{$^{2}$ Bristol Centre for Complexity Sciences, School of Chemistry, University of Bristol, Bristol, BS8
1TS, UK}

\address{$^{3}$ Research School of Chemistry, The Australian National University,
Canberra, ACT 0200, Australia}

\ead{paddy.royall@bristol.ac.uk}

\address{Received September 30th, 2009}
\begin{abstract}
We study the structure of colloidal fluids with reference to colloid-polymer mixtures.
We compare the one component description of the Asakura-Oosawa (AO) idealisation of colloid-polymer mixtures with the full two-component
model. We also consider the Morse potential, a variable range interaction, for which the ground state clusters are known.
Mapping the state points between these systems, we find that the pair structure of the full AO model is equally well described by the Morse potential or the one component AO approach. We employ a recently developed method to identify in the bulk fluid the ground state clusters 
relevant to the Morse potential. Surprisingly, when we measure 
the cluster populations, we find that the Morse fluid is significantly closer the full AO fluid than the one component AO description.
\end{abstract}


\section{Introduction}

\label{secIntroduction}

Although in principle colloidal dispersions are rather complex multicomponent
systems, the spatial and dynamic asymmetry between the colloidal particles
(10 nm-1 $\mu$m) and smaller molecular and ionic species has led
to schemes where the smaller components are formally integrated out
\cite{likos2001}. This leads to an effective one-component picture, where only
the \emph{effective} colloid-colloid interactions need be considered.
The behaviour in the original complex system may then be faithfully
reproduced by appealing to liquid state theory~\cite{hansen} and
computer simulation \cite{frenkel}. Since the shape of the particles
is typically spherical, and the effective colloid-colloid interactions
may be tuned, it is often possible to use models of simple liquids
to accurately describe colloidal dispersions. 

Central to this one-component approach is the use of a suitable colloid-colloid
interaction $u(r)$. Notable early successes include the Derjaguin,
Landau, Verwey and Overbeek theory of charged colloids~\cite{overbeek}
and the Asakura-Oosawa (AO) theory of colloids in a solution of polymers~\cite{asakura1954, asakura1958},
subsequently popularised by Vrij~\cite{vrij1976}. While theories
such as these have been used to describe colloidal model systems in
which the interactions may be tailored with very considerable success~\cite{pusey1986,monovoukis1989,poon2002},
the general situation is often considerably more complex. 

In the colloid-polymer mixtures of interest here, the effective colloid-colloid
interactions are set by the polymer chemical potential. One imagines a polymer reservoir 
coupled to the colloidal suspension, in which, if the polymers are ideal as assumed by AO, then
the polymer chemical potential is proportional to the concentration.
In practice, experimental systems seldom feature coupled polymer reservoirs, so
one is often limited to knowledge of the polymer concentration 
in the sample cell; for a given polymer concentration, the chemical potential varies with 
colloid volume fraction, due to the volume excluded to the polymer by the colloids.
The volume accessible to polymer is also dependent upon phase separation and colloidal crystallisation.
In other words, the effective colloid-colloid interaction can vary with colloid concentration
and also change as a function of time, giving rise to novel kinetic pathways and (unlike simple atomic substances), 
a triple coexistence \emph{region} \cite{poon1999}; meanwhile external fields such as gravity may 
couple with the multi-component nature of the colloid-polymer system to yield novel phenomena such as floating 
colloidal liquids \cite{schmidt2004}.

Even in the case of a one-phase colloidal fluid in coexistence with a polymer reservoir,
for polymer-colloid size ratio $q>0.154$ \cite{dijkstra1999}, the effective colloid-colloid interaction 
has a many-body component and thus
is dependent upon colloid volume fraction,while for smaller 
size ratios the one-component mapping has been shown to be exact \cite{dijkstra2000}.
Nevertheless, one may
integrate out the polymer degrees of freedom to arrive at an effective one-component description
for the colloids, as given by AO \cite{asakura1958} and Vrij \cite{vrij1976}. It is worth noting that there is more than one approach
to determining the effective one-component interaction in a multicomponent system, and that these
do not always give the same result \cite{louis2002b}. The effective one-component description has since 
been extended to include these many-body effects \cite{moncho2003,dijkstra2002s,dijkstra2006,vink2005s}. 
Other important departures from the assumptions of Asakura and Oosawa include
non-ideal polymer-polymer interactions \cite{louis2002b,bolhius2002}, which
have considerable implications for phase behaviour and interfacial properties \cite{fortini2008i}
along with electrostatic interactions between the colloids \cite{fortini2005}.

The validity of the one-component approach in describing the colloid-colloid interactions has also been investigated experimentally.
The interaction between
a colloid and a glass wall can be accurately measured with total internal
reflection microscopy~\cite{bechinger1999}, while the interaction
between two colloids confined to a line can be measured using optical
tweezers~\cite{crocker1994,verma1998}. 
An alternative approach is to measure correlation functions and invert
them to extract the effective potentials. Traditionally this has been
achieved by scattering techniques that measure the reciprocal space
structure factor $S(k)$\cite{hansen,ye1996}.
Another means is to determine the structure in real space 
in 2D and 3D at the single
particle level using optical microscopy 
\cite{royall2003,brunner2002}, after making some assumptions about the system,
one may deduce the effective colloid-colloid potential. 
This may be done with sufficient precision
that interaction potentials can be quite accurately determined both for
purely repulsive systems \cite{brunner2002,royall2006} and for systems
with attractive interactions \cite{royall2007}. 

\begin{figure}
\begin{centering}
\includegraphics[width=7cm]{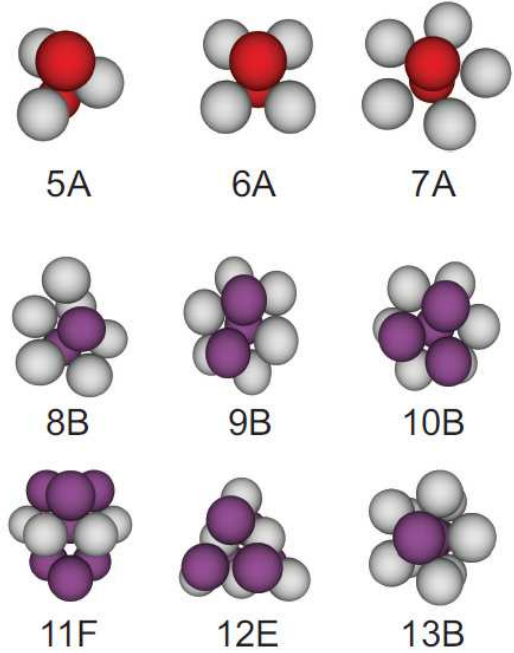} 
\par\end{centering}
\caption{(color online) The ground state clusters for 
the short ranged Morse potential ($\rho_0=25.0$) for $m < 14$ particles.
Here we follow the nomenclature of Doye \emph{et. al.}  \cite{doye1995}. }
\label{figTCC} 
\end{figure}

The possibility of direct visualisation of colloidal fluids also allows, for example the clusters formed
to be studied \cite{sedgwick2004,lu2008,ohtsuka2008}. Lu \emph{et. al.} explored the idea, introduced by
Noro and Frenkel in their `extended law of corresponding states' 
\cite{noro2000}, that the structure of these dilute attractive fluids
(the so-called energetic fluid regime \cite{louis2001}) is somewhat 
insensitive to the exact nature of the potential \cite{lu2008}. We have also recently argued that the (known)
ground state clusters formed by systems interacting under the Morse potential (fig. \ref{figTCC}) are
also relevant to colloid-polymer mixtures. Interestingly, recent work suggests that
in fact, hard core systems such as colloid-polymer mixtures might exhibit
somewhat richer (degenerate) topologies of ground state clusters, as more than one structure can have
identical numbers of bonds \cite{arkus2009}.

Here, we investigate the validity of the one-component approach in colloid-polymer mixtures
by comparing the full Asakura-Oosawa multi-component model with explicit polymers and the one-component AO model \cite{asakura1958, vrij1976}. 
Given a suitable choice of parameters, the variable-ranged Morse potential can provide a
good approximation to the one-component AO potential.
In addition to the fact that the ground state clusters are known for the Morse potential, we note that its
continuous form is amenable to Brownian and molecular dynamics computer simulations.
We therefore also compare the Morse potential by applying
the law of corresponding states to map the Morse to the one component AO
interaction. 
We consider the structure of the resulting dilute colloidal fluids. In addition to conventional pair-correlation function-based
methods, we employ a recent-developed method which identifies structures topologically equivalent
to isolated clusters \cite{williams2007}. 

This paper is organised as follows. In section \ref{secSimulations-and-Interaction}
we introduce the simulation methodology and our approach for comparing
different interaction potentials, our results are presented in section
\ref{secResults} and we conclude with a discussion in section \ref{secConclusions}.

\section{Simulations and Interaction Potentials}

\label{secSimulations-and-Interaction}

The seminal theory of colloid-polymer mixtures is that of Asakura
and Oosawa~\cite{asakura1954, asakura1958}. Here colloids are treated as hard spheres with no permitted overlap.
Polymers are ideal, and may freely overlap with one another, but the polymer-colloid interaction is 
also hard, in that no overlap is permitted. That is to say, the colloid-colloid interaction $u_{CC}$, 
colloid-polymer interaction $u_{CP}$ and polymer-polymer interaction $u_{PP}$ read

\begin{eqnarray}  
\beta u_{CC}(r)=
\cases{\infty&for $r \le \sigma$\\ 
       0&for $r > \sigma$\\
       }\nonumber \\
\beta u_{CP}(r)=       
\cases{\infty&for $r \le (\sigma+\sigma_P)/2$\\ 
       0&for $r > (\sigma+\sigma_P)/2$\\
       }\label{eqUAOtrue} \\
\beta u_{PP}(r)=0.\nonumber
\end{eqnarray}

\noindent where $r$ is the centre to centre separation of the two
colloids/polymers and $\beta=1/k_{B}T$, where $T$ is temperature, $k_B$ is Boltzmann's constant. 
$\sigma$ and $\sigma_P$ are the diameters of the colloids and polymers respectively. 

Some comments on the derivation of the one-component description are in order. For a more 
complete description the reader is referred to Dijkstra \emph{et. al.} \cite{dijkstra1999}. The Hamiltonian
of the AO model is thus


\begin{equation} 
H=H_{CC}+H_{CP}+H_{PP} \\
\label{eqAOHamiltonian}
\end{equation}

\noindent where

\begin{eqnarray} 
H_{CC}=\sum^{N_C}u_{CC}(r)  \\
H_{CP}=\sum^{N_C}\sum^{N_P}u_{CP}(r)  \\
H_{PP}=\sum^{N_P}u_{PP}(r)=0   \label{eqAOHamiltonianBits}
\end{eqnarray}

\noindent where $N_C$ and $N_P$ are the respective numbers of colloids and polymers. Dijkstra \emph{et. al.}
cast the thermodynamic potential $F$ of the colloid-polymer system as 

\begin{eqnarray} 
\exp[-\beta F] &= \sum^{\infty}_{N_P=0} \frac{z^{N_P}_P}{N_C!\Lambda_C^{3N_C}N_P!}   \int_VdR^{N_C}   \int_VdR^{N_P} \exp[-\beta (H_{CC}+H_{CP})] \\
&= \frac{1}{N_C!\Lambda_C^{3N_C} } \int_VdR^{N_C}\exp[-\beta H^{EFF}]
\label{eqEffectiveHamiltonian}
\end{eqnarray}

\noindent where $z_P$ is the polymer fugacity $\Lambda_C$, is the thermal De Broglie wavelength of the colloids, $R^{N_C}$ and $R^{N_P}$ are the coordinates of the colloids and polymers respectively. $H^{EFF}=N_{CC}+\Omega$ is the \emph{effective Hamiltonian} of the colloids. 

Now $\Omega$ is the grand potential of the fluid of ideal polymer coils in an
external field of $N_C$ colloids with coordinates $R$, and may be expanded as

\begin{equation} 
\Omega = \Omega_0 + \Omega_1 + \Omega_2 + \dots  
\label{eqOmega}
\end{equation}

\noindent where $\Omega_0$ is a 0-body term (the Grand potential of an ideal polymer system) $\Omega_1$ is a 1-body term related to the volume excluded by the $N_C$ colloids and $\Omega_2$ is the two-body term. Dijkstra \emph{et. al.}
show that all higher order terms are zero for polymer colloid size ratios $q=\sigma_P/\sigma<0.154$ \cite{dijkstra2000}. The two-body term

\begin{equation} 
\Omega_2 = \sum_{N_C} \beta u_{AO}(r) 
\label{eqAOOmega2}
\end{equation}

\noindent where

\begin{equation} 
\beta u_{AO}(r)=
\cases{
\frac{\pi(2R_{G})^{3}z_{P}}{6}\frac{(1+q)^{3}}{q^{3}}&\\ \times\{1-\frac{3r}{2(1+q)\sigma}+\frac{r^{3}}{2(1+q)^{3}\sigma^{3}}\}&for $\sigma< r\le\sigma+(2R_{G}),$\\ 0&for $r > \sigma+(2R_{G}),$\\
} 
\label{eqAO}
\end{equation}

%

\noindent Now the polymer fugacity $z_{P}$
is equal to the number density $\rho_{PR}$ of ideal polymers in a
reservoir at the same chemical potential as the colloid-polymer mixture.
Thus within the AO model, the effective temperature is inversely proportional
to the polymer reservoir concentration. The interaction induced by the polymers in
equation (\ref{eqAO}) is identical to that given by AO \cite{asakura1958} and Vrij \cite{vrij1976}.

We also use the Morse potential which reads

\begin{equation}
\beta u_{M}(r)=\beta \varepsilon_{M}\exp[\rho_{0}(\sigma-r)]\{\exp[\rho_{0}(\sigma-r)]-2\}
\label{eqMorse}
\end{equation}

\noindent where $\rho_{0}$ is a range parameter and $\beta\varepsilon_{M}$
is the potential well depth. 
We set $\rho_0=25.0$ to simulate a system with short-ranged attractions similar to a colloid-polymer mixture.

\begin{figure}
\begin{centering}
\includegraphics[width=7.0cm]{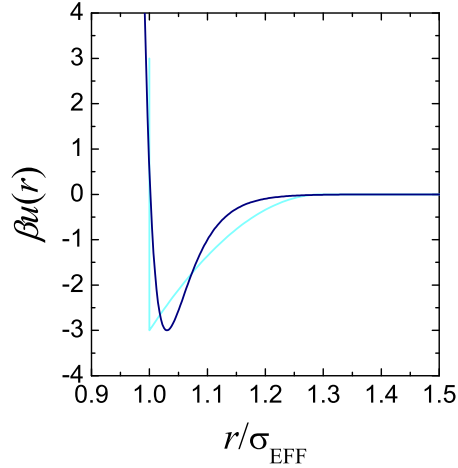} 
\par\end{centering}
\caption{(color online) Interaction potentials used: Morse
(blue) and one-component Asakura-Oosawa (cyan). Both are scaled by the effective
hard sphere diameter $\sigma_{EFF}$.}
\label{figU}
\end{figure}

\subsection{Comparing different systems}

In order to match state points
between the Morse and one component Asakura-Oosawa interactions, we use the extended
law of corresponding states introduced by Noro and Frenkel \cite{noro2000}.
Specifically, this requires two interactions to have identical well depths and reduced
second virial coefficients $B_{2}^{*}$ where 

\begin{equation}
B_{2}^{*}=B_{2}/\frac{2}{3}\pi\sigma_{EFF}^{3}
\label{eqReducedVirial}
\end{equation}

\noindent where $\sigma_{EFF}$ is the effective hard sphere diameter and the
second virial coefficient

\begin{equation}
B_{2}= 2 \pi \intop^{\infty}_0 drr^{2}\left[1-\exp\left(-\beta u(r)\right)\right].
\label{eqCorrespondingStates}
\end{equation}

The effective hard sphere diameter is defined as

\begin{equation}
\sigma_{EFF}=\intop^{\infty}_0 dr\left[1-\exp\left(-\beta u_{REP}(r)\right)\right]
\label{eqEffective}
\end{equation}

\noindent where the repulsive part of the \emph{potential}
$u_{REP}$ is where $u(r)>0$. Thus we compare different interactions by equating $B_{2}^{*}$ and
$\sigma_{EFF}$. The latter condition leads to a constraint on number density
\begin{equation}
\rho_{EFF}=\frac{N\pi\sigma_{EFF}^{3}}{6V}
\label{eqDensity}
\end{equation}

\noindent where $V$ is the volume of the simulation box.

\subsection{Simulation Details}

For the one-component systems, we use standard Monte-Carlo (MC) simulation in the NVT ensemble \cite{frenkel} with 
$N=2048$ particles. Each simulation was typically equilibrated for
$10^{7}$ MC moves and run for a further $10^{7}$ moves.
For each state point we performed ten independent simulation runs.
We confirmed that the system was in equilibrium on the
simulation timescale by monitoring the potential energy. The Morse potential is truncated and shifted at $r=2.5\sigma$.
In the case of the full AO system, we use Monte-Carlo simulation, with polymers included
grand-canonically \cite{frenkel,vink2004}. The interaction potential for the one-component AO is taken as
eq. (\ref{eqAO}) with the additional hard sphere colloid-colloid interaction $u_{CC}(r)$ [eq. (\ref{eqAOHamiltonianBits})].

We match the Morse and one component AO using eqs. (\ref{eqReducedVirial}) (\ref{eqDensity}) by requiring the interactions to have the same well depth We set a well depth of $2.0k_BT$ and colloid volume fractions of $\phi_C=\pi\sigma^3\rho_C/6=0.05$, $\phi_C=0.25$ and $\phi_C = 0.445$, where $\rho_C$ is the colloid number density. For the Morse interaction, with 
range parameter $\rho_0=25.0$, this leads to an effective hard sphere diameter $\sigma_{EFF}\approx0.9696\sigma$ [eq. (\ref{eqEffective})]. Applying equation (\ref{eqDensity}) we therefore have a slightly higher volume fraction in the Morse system of $\phi_M\approx1.1097\phi_C$. In the one-component AO system, 
these Morse parameters map via eq. (\ref{eqAO}) and eq. (\ref{eqReducedVirial}) to a polymer-colloid size ratio of $q \approx 0.2575$ and polymer reservoir number density $\rho_{PR} \approx 0.5597 \sigma_P^{-3}$.
It is worth noting that there is some sensitivity in the mapping we have used to the depth of the attractive well. We 
have taken a value of $\beta \epsilon_{M}=2.0$, which we fix throughout this work. However, the `hardness' of the Morse potential depends upon $ \beta \epsilon_{M} $, as, consequently, does the effective hard sphere diameter. In principle, one should therefore repeat the mapping for each $\beta \epsilon_{M}$.

The full AO system is challenging to simulate, especially when there is a considerable size discrepancy between the colloids and polymers,
leading to very large numbers of particles in the system
 \cite{dijkstra2006}. Of course, this is one of the attractions of using a one component description. Here we could only equilibrate the system 
 to our satisfaction for the higher densities, $\phi_C=0.25$ and $\phi_C=0.445$, owing to the vastly reduced number of polymers at higher colloid density. We used $N=256$ and $N=512$ for $\phi_C=0.25$ and $\phi_C=0.445$ respectively. The system was equilibrated for $3\times 10^7$ MC moves of either polymer or colloid in each case. Unlike the one component systems, two simulations per state point were performed in the case of the full AO system.
In comparing the full AO system with the one component systems, we only consider the colloids and ignore the polymer coordinate data.


\subsection{The topological cluster classification}

To analyse the structure, we identify the bond network using the Voronoi
construction. Having identified the bond network, we use the Topological
Cluster Classification (TCC) to determine the nature of the clusters in the bulk fluid~\cite{williams2007}.
This analysis identifies all the shortest path three, four and five
membered rings in the bond network. We use the TCC to find clusters
which are global energy minima of the Morse potential for $\rho_0=25.0$.
These clusters are shown in figure \ref{figTCC}.
We identify all topologically distinct Morse Clusters.
In addition, for $m=13$ clusters
we identify the FCC and HCP thirteen particle structures in terms
of a central particle and its twelve nearest neighbours. We illustrate
these clusters in fig. \ref{figTCC}. For more details see \cite{williams2007}.
We found relatively little clustering at the moderate attractions $\beta \varepsilon=2.0$ 
at lower and intermediate densities, thus we present TCC
results for the highest density studied, $\phi_C=0.445$.

\section{Results and Discussion}

\label{secResults}

We begin our presentation of the results by comparing the pair correlation
functions of the various systems at differing densities, followed by the TCC analysis. 

\begin{figure}
\begin{centering}
\includegraphics[width=6.0cm]{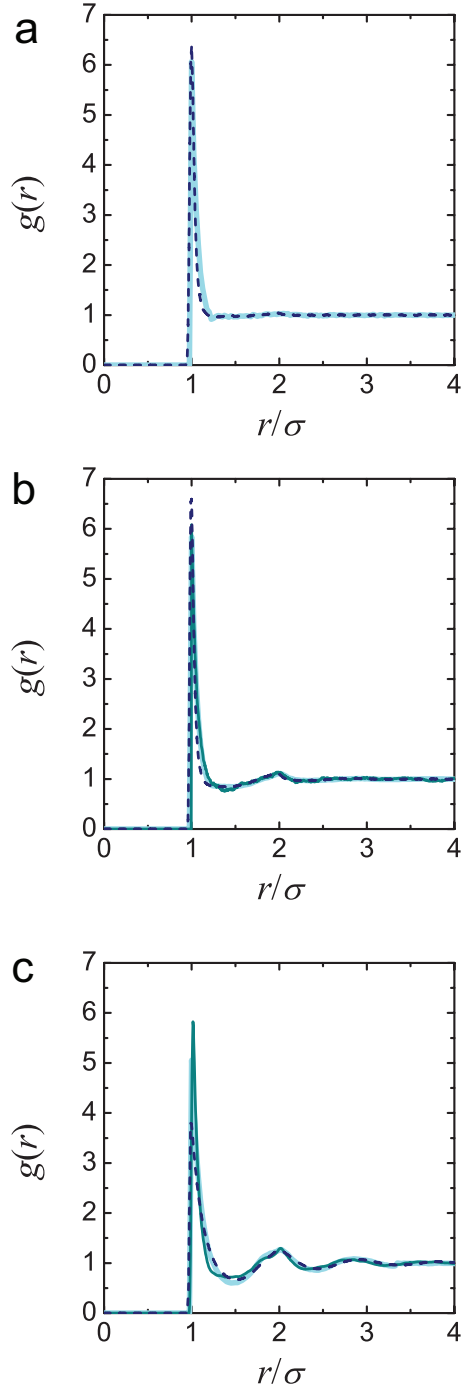} 
\par\end{centering}

\caption{(color online) Pair-correlation functions at various densities. (a) Low density, $\phi_C=0.05$,
(b) moderate density ($\phi_C=0.25$) and (c) high density ($\phi_C=0.445$). The $g(r)$ for the Morse 
potential is shown in dark blue (dashed), the full AO system in turquoise and the pale blue is the one-component AO.}
\label{figG} 
\end{figure}

Pair correlation functions are shown in figure \ref{figG}. At low density, $g(r) \approx \exp [ -\beta u(r) ]$. This is illustrated in 
both cases in figure \ref{figG}(a) ($\phi_C=0.05$), in the form of a strong peak at contact, reflecting the short-ranged nature of these attractions.
There are some minor differences. These are in general consistent with the differences obtained from the potentials, figure \ref{figU}, upon taking the low density limit, $g(r)\approx \exp[-\beta u(r)]$. For example, the slighter softer Morse potential leads to a slightly slower decay at 
$r<\sigma$. Likewise, in the range $1.1 \sigma \leq r \leq 1.2 \sigma$, the AO decays to unity rather slower than the Morse, reflecting 
the greater magnitude of the AO in that range. In general, however, the agreement between the Morse and AO systems is good. 

We now turn to higher densities, in particular to $\phi_C=0.25$ [fig. \ref{figG}(b)]. In this case, we were able to equilibrate the full AO system in addition to the one-component descriptions.
Packing leads to a second peak around $2\sigma$. Again, we see a similar behaviour between the different systems. Significantly, the small differences between the $g(r)$s, comparing Morse to firstly the one component AO and then the full AO, are similar. That is to say, the one-component AO, which, for example does not include many-body interactions \cite{moncho2003,dijkstra2006}, shows discrepancies comparable to the Morse potential in its description of the full AO system. 

At the highest density studied ($\phi_C=0.445$), overall we find a similar behaviour, as may be seen in figure \ref{figG}(c). This is not altogether surprising,
as in dense liquids, the structure is well-known to be largely dominated by the hard core \cite{barker1976}. Some differences are, however apparent.
The Morse system has a weaker first peak, than either the one component or full AO systems. This is likely due to the lack of an infinitely hard core in the Morse interaction. The first peak notwithstanding, the differences between all three systems are comparable. In comparing the one-component AO and full AO, our results are compatible with the results of Dijkstra \emph{et. al.}, who found that $g(r)$s 
produced from the two descriptions were indistinguishable in the case of $q=0.15$ where the one-component description is exact \cite{dijkstra2000}.

\begin{figure}
\begin{centering}
\includegraphics[width=9.0cm]{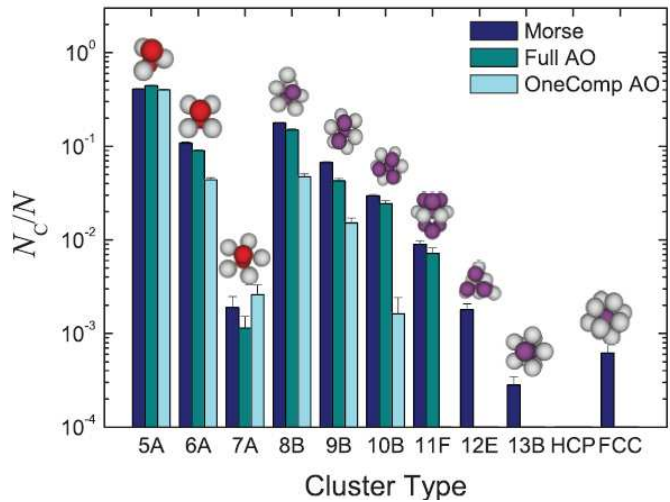} 
\par\end{centering}
\caption{(color online) (color online) Population of particles in a given cluster, for $\phi_C=0.445$. 
$N_c$ is the number of particles in a given cluster,
$N$ the total number of particles sampled. Here we consider only ground state clusters for the Morse $\rho_0=25.0$ system. 
Dark blue denotes Morse, turquoise the full AO and light blue the one component AO.}
\label{figClusGround} 
\end{figure}

We now turn our attention to the cluster populations in the dense system (fig. \ref{figClusGround}). 
In all these systems, a \emph{range} of different clusters are found, with none dominating.
Thus we argue, that when considering energetically locally favoured structures (i.e. clusters),
it is important to consider the possibility that more than one topology may be important.
The overall behaviour between the systems is similar. Among the more populous, smaller clusters, the 7A pentagonal bipyramid
has a rather low population. However  7A is also found as part of 
larger clusters, notably 8B. According to our counting algorithm, if a given particle is part of both a 7A and 8B
cluster, it is taken as 8B only.
A few particles are found as FCC crystal fragments (we found no HCP type environments). 

In comparing these systems we see that the one component AO forms rather fewer clusters for $8 \leq m \leq 10$ than the other systems, and none
at higher $m$. Our statistics are necessarily more limited for the full AO system, which we believe restricts our ability to determine the population of rarer, higher order clusters. For $m \leq 11$, the \emph{Morse} and full AO have rather similar populations, 
except that the cluster population in slightly higher for the Morse system in the case that $m \ge 6$.
We thus argue that in this respect the Morse potential accurately reproduces the full AO model.

\section{Discussion and Conclusions}

\label{secConclusions}

We have analysed the pair structure and performed a topological cluster classification on a range of model systems for
colloid-polymer mixtures. Using the extended law of corresponding states \cite{noro2000}, we have mapped the variable ranged Morse potential to a well-known one component model for AO colloid-polymer mixtures. We have also considered the full Asakura-Oosawa model.
In general, we find good agreement between all three systems. The relatively small difference in the pair structure between the slightly soft Morse potential and one component AO system seems to be accounted for by noting the differences in their functional form (fig. \ref{figU}).
The small discrepancies exhibited between the full AO and the one component systems favour either. That is to say, our $g(r)$ results suggest that
the Morse potential does \emph{as good} a job of describing the full AO system as the one component AO system.

Although the pair structure may be very similar between these three systems, the topological cluster classification reveals significant differences.
In particular, the one component AO system forms fewer higher order clusters for $m \ge 8$ (8B clusters alone account for 20\% of the particles in the other systems) and we detect no clusters at all for $m \ge 10$. In this respect, \emph{the Morse potential does a better job than the one component Asakura-Oosawa interaction in describing the full AO system}.

Some pointers for further work are considered. Dijkstra \emph{et. al.} \cite{dijkstra2002s, dijkstra2006} have developed an elegant means by which the many-body effects implicit in the full AO model are taken into account. It would be most attractive to subject this system 
to an analysis similar to that presented here. 
Recalling that we were unable to obtain sufficient statistics to
calculate a $g(r)$ for the full AO system for $\phi_C=0.05$, we note that accelerated MC methods such as the cluster move of Vink and Horbach \cite{vink2004} would be most helpful in generating sufficient statistics. 

Moving closer to experiments, non-ideal polymers \cite{bolhius2002} and electrostatic interactions \cite{fortini2005} may all impact on these conclusions. We have also considered only a few state points. Furthermore, we have neglected polydispersity, omnipresent in experimental colloidal systems, which has the potential to alter the results of an analysis similar to that carried out here. Coordinate tracking, particularly in 3D experiments based around confocal microscopy, \emph{is prone to measurement errors of around $0.02-0.05\sigma$} \cite{royall2007}. Work to investigate the sensitivity of this analysis to such experimental considerations is in progress. Early indications are that the TCC analysis is surprisingly robust to experimental tracking errors and polydispersity.

The system we have chosen (probably) does not have a stable gas-liquid-coexistence. 
However $q=0.2575$ is somewhat above the value of $q=0.154$ at which 3-body and higher order interactions
vanish in the AO model \cite{dijkstra1999, dijkstra2000}; these effects may be non-negligible but the similarity in the correlation functions we measure suggests that the effects to not too large, although larger polymers would lead to stronger many-body effects \cite{dijkstra2006}. Furthermore, 
larger polymers lead to such a coexistence between colloidal `gas' and `liquid'. The location of the critical point is known to be strongly dependent upon the exact model chosen \cite{vink2004, loVerso2006}. Moving closer to the critical point, we expect to find different results upon comparing the various models.

Finally, we have considered equilibrium fluids. The behaviour out of equilibrium is most important, particularly in the case of, for example colloidal gels \cite{royall2008gel}. However, we are unaware of suitable simulation models for non-equilibrium studies, except one-component descriptions
with softened cores \cite{puertas2004, fortini2008}, and the Morse potential \cite{royall2008gel}. It is almost necessary to use one-component descriptions out of equilibrium, due to the degree of computation required. Moreover, Brownian dynamics, appropriate to out-of-equilibrium situations,
is challenging to implement with hard interactions. Out of equilibrium, hydrodynamic interactions may also play a role, and have recently been applied to attractive colloidal systems \cite{monchoJorda2009}.

\section*{Acknowledgements}

JT and CPR thank the Royal Society for funding, AM acknowledges the support of EPSRC grant EP/5011214. The authors are grateful to M. Caine, D. Klotsa and R. Jack for helpful discussions.

\section*{References}


\end{document}